\documentclass[prl,twocolumn,aps,superscriptaddress]{revtex4}
\usepackage{color}
\usepackage{hyperref}
\usepackage{graphicx}
\usepackage{amsmath}
\usepackage{bbold}
\usepackage{slashed}

\usepackage{amssymb}
\begin{document}

%

%\begin{document}

%\newcommand{\Ga}{\mathcal{G}}
\newcommand{\Gb}{\pazocal{G}}
\newcommand{\be}{\begin{equation}}
\newcommand{\ee}{\end{equation}\noindent}
\newcommand{\bear}{\begin{eqnarray}}
\newcommand{\ear}{\end{eqnarray}\noindent}
\newcommand{\no}{\noindent}
\newcommand{\non}{\nonumber\\}

%reference to equations
\def\veps#1{\varepsilon_{#1}}
\def\ddel{{}^\bullet\! \Delta}
\def\deld{\Delta^{\hskip -.5mm \bullet}}
\def\dddel{{}^{\bullet \bullet} \! \Delta}
\def\ddeld{{}^{\bullet}\! \Delta^{\hskip -.5mm \bullet}}
\def\deldd{\Delta^{\hskip -.5mm \bullet \bullet}}
\def\epsk#1#2{\varepsilon_{#1}\cdot k_{#2}}
\def\epseps#1#2{\varepsilon_{#1}\cdot\varepsilon_{#2}}
\def\eq#1{{eq. (\ref{#1})}}
\def\eqs#1#2{{eqs. (\ref{#1}) -- (\ref{#2})}}
\def\t#1{\tau_1}
\def\mn{{\mu\nu}}
\def\rs{{\rho\sigma}}
\newcommand{\Det}{{\rm Det}}
\def\Tr{{\rm Tr}\,}
\def\tr{{\rm tr}\,}
\def\sumij{\sum_{i<j}}
\def\e{\,{\rm e}}
\def\eps{\varepsilon}
%r.schuetzhold@hzdr.de

\def\bddel{{}^\bullet\! {\underline\Delta}}
\def\bdeld{{\underline\Delta}^{\hskip -.5mm \bullet}}
\def\bdddel{{}^{\bullet \bullet} \! {\underline\Delta}}
\def\bddeld{{}^{\bullet}\! {\underline\Delta}^{\hskip -.5mm \bullet}}
\def\bdeldd{{\underline\Delta}^{\hskip -.5mm \bullet \bullet}}

%worldline

\def\fr#1#2{\frac{#1}{#2}}
\def\half{\frac{1}{2}}
\def\freeexp{{\rm e}^{-\int_0^Td\tau {1\over 4}\dot x^2}}
\def\kinb{{1\over 4}\dot x^2}
\def\kinf{{1\over 2}\psi\dot\psi}
\def\expk{{\rm exp}\biggl[\,\sum_{i<j=1}^4 G_{Bij}p_i\cdot p_j\biggr]}
\def\expp{{\rm exp}\biggl[\,\sum_{i<j=1}^4 G_{Bij}p_i\cdot p_j\biggr]}
\def\expshort{{\e}^{\half G_{Bij}p_i\cdot p_j}}
\def\expabb{{\e}^{(\cdot )}}
\def\epseps#1#2{\varepsilon_{#1}\cdot \varepsilon_{#2}}
\def\epsk#1#2{\varepsilon_{#1}\cdot k_{#2}}
\def\epsr#1#2{r_{#2}\cdot\varepsilon_{#1}}
\def\rk#1#2{r_{#1}\cdot p_{#2}}
\def\kk#1#2{k_{#1}\cdot k_{#2}}
\def\G#1#2{G_{B#1#2}}
\def\Gp#1#2{{\dot G_{B#1#2}}}
\def\GF#1#2{G_{F#1#2}}
\def\Dab{{(x_a-x_b)}}
\def\Dsq{{({(x_a-x_b)}^2)}}
\def\PITD{{(4\pi T)}^{-{D\over 2}}}
\def\4piTD{{(4\pi T)}^{-{D\over 2}}}
\def\4piT4{{(4\pi T)}^{-2}}
\def\TintmD{{\dps\int_{0}^{\infty}}{dT\over T}\,e^{-m^2T}
    {(4\pi T)}^{-{D\over 2}}}
\def\Tintm4{{\dps\int_{0}^{\infty}}{dT\over T}\,e^{-m^2T}
    {(4\pi T)}^{-2}}
\def\Tintm{{\dps\int_{0}^{\infty}}{dT\over T}\,e^{-m^2T}}
\def\Tint{{\dps\int_{0}^{\infty}}{dT\over T}}
\def\np{n_{+}}
\def\nm{n_{-}}
\def\Np{N_{+}}
\def\Nm{N_{-}}
\def\ed{e^{(\cdot)}}
\def\t#1{\tau_{#1}}
\def\et#1#2{e^{({#1}\rightarrow{#2})}}
\def\ett#1#2#3#4{e^{({#1}\rightarrow{#2}, #3\rightarrow#4)}}
\newcommand{\slG}{{{\dot G}\!\!\!\! \raise.15ex\hbox {/}}}
\newcommand{\Gd}{{\dot G}}
\newcommand{\Gund}{{\underline{\dot G}}}
\newcommand{\Gdd}{{\ddot G}}
\def\GBd12{{\dot G}_{B12}}
\def\Dx{\dps\int{\cal D}x}
\def\Dy{\dps\int{\cal D}y}
\def\Dpsi{\dps\int{\cal D}\psi}
\def\dint#1{\int\!\!\!\!\!\int\limits_{\!\!#1}}
\def\ddtau{{d\over d\tau}}
\def\ie{\hbox{$\rm style{\int_1}$}}
\def\iz{\hbox{$\rm style{\int_2}$}}
\def\id{\hbox{$\rm style{\int_3}$}}
\def\ldop{\hbox{$\lbrace\mskip -4.5mu\mid$}}
\def\rdop{\hbox{$\mid\mskip -4.3mu\rbrace$}}
\def\bdel{{\underline\Delta}}
%
%VARIOUS
\newcommand{\1}{{\'\i}}
\def\dps{\displaystyle}
\def\sy{\scriptscriptstyle}
\def\sy{\scriptscriptstyle}

\def\del{\partial}
\def\deli{\partial_{\kappa}}
\def\delj{\partial_{\lambda}}
\def\delk{\partial_{\mu}}
\def\delij{\partial_{\kappa\lambda}}
\def\delik{\partial_{\kappa\mu}}
\def\deljk{\partial_{\lambda\mu}}
\def\delki{\partial_{\mu\kappa}}
\def\delkl{\partial_{\mu\nu}}
\def\delijk{\partial_{\kappa\lambda\mu}}
\def\deljkl{\partial_{\lambda\mu\nu}}
\def\delikl{\partial_{\kappa\mu\nu}}
\def\delijkl{\partial_{\kappa\lambda\mu\nu}}
\def\delijklm{\partial_{\kappa\lambda\mu\nu o}}
\def\O(#1){O($T^#1$)} 
\def\O2{O($T^2$)}
\def\O3{O($T^3$)}
\def\O4{O($T^4)}
\def\O5{O($T^5$)}
\def\dA{\partial^2}
\def\DA{\sqsubset\!\!\!\!\sqsupset}
\def\eins{  1\!{\rm l}  }
\def\a#1{\alpha_{#1}}
\def\b#1{\beta_{#1}}
\def\m#1{\mu_{#1}}
\def\n#1{\nu_{#1}}
\def\m#1{\mu_{#1}}
\def\n#1{\nu_{#1}}
\def\a{\alpha}
\def\b{\beta}
\def\m{\mu}
\def\n{\nu}
\def\s{\sigma}
\def\r{\rho}
\def\e{{\rm e}}
\def\z{\zeta}
\def\vareps{\varepsilon}

\def\gF{\gamma_{\mathcal{F}}}
\def\gG{\gamma_{\mathcal{G}}}
\def\gFF{\gamma_{\mathcal{F}\mathcal{F}}}
\def\gGG{\gamma_{\mathcal{G}\mathcal{G}}}
\def\gFG{\gamma_{\mathcal{F}\mathcal{G}}}

\newcommand{\Vka}{V_{\kappa}}
\newcommand{\Vla}{V_{\lambda}}
\newcommand{\Vmu}{V_{\mu}}
\newcommand{\Vnu}{V_{\nu}}
\newcommand{\Vro}{V_{\rho}}
\newcommand{\Vkala}{V_{\kappa\lambda}}
\newcommand{\Vkamu}{V_{\kappa\mu}}
\newcommand{\Vkanu}{V_{\kappa\nu}}
\newcommand{\Vlamu}{V_{\lambda\mu}}
\newcommand{\Vlanu}{V_{\lambda\nu}}
\newcommand{\Vlaka}{V_{\lambda\kappa}}
\newcommand{\Vmunu}{V_{\mu\nu}}
\newcommand{\Vmuka}{V_{\mu\kappa}}
\newcommand{\Vnuro}{V_{\nu\rho}}
\newcommand{\Vkalamu}{V_{\kappa\lambda\mu}}
\newcommand{\Vkalanu}{V_{\kappa\lambda\nu}}
\newcommand{\Vkalaro}{V_{\kappa\lambda\rho}}
\newcommand{\Vkamunu}{V_{\kappa\mu\nu}}
\newcommand{\Vlamunu}{V_{\lambda\mu\nu}}
\newcommand{\Vmunuro}{V_{\mu\nu\rho}}
\newcommand{\Vkalamunu}{V_{\kappa\lambda\mu\nu}}
\newcommand{\Fkala}{F_{\kappa\lambda}}
\newcommand{\Fkanu}{F_{\kappa\nu}}
\newcommand{\Flaka}{F_{\lambda\kappa}}
\newcommand{\Flamu}{F_{\lambda\mu}}
\newcommand{\Fmunu}{F_{\mu\nu}}
\newcommand{\Fnumu}{F_{\nu\mu}}
\newcommand{\Fnuka}{F_{\nu\kappa}}
\newcommand{\Fmuka}{F_{\mu\kappa}}
\newcommand{\Fkalamu}{F_{\kappa\lambda\mu}}
\newcommand{\Flamunu}{F_{\lambda\mu\nu}}
\newcommand{\Flanumu}{F_{\lambda\nu\mu}}
\newcommand{\Fkamula}{F_{\kappa\mu\lambda}}
\newcommand{\Fkanumu}{F_{\kappa\nu\mu}}
\newcommand{\Fmulaka}{F_{\mu\lambda\kappa}}
\newcommand{\Fmulanu}{F_{\mu\lambda\nu}}
\newcommand{\Fmunuka}{F_{\mu\nu\kappa}}
\newcommand{\Fkalamunu}{F_{\kappa\lambda\mu\nu}}
\newcommand{\Flakanumu}{F_{\lambda\kappa\nu\mu}}

% commands
\newcommand{\tvec}{\vec}
\newcommand{\action}{\mathscr{S}}
\newcommand{\pathdiff}[1]{\!\mathscr{D}#1\,}
\newcommand{\diff}[1]{\!\mathrm{d}#1\,}
\newcommand{\dottau}{\accentset{\boldsymbol\circ}}
\newcommand{\dott}{\accentset{\mbox{\large .}}}
\newcommand{\ddott}{\accentset{\mbox{\large ..}}}
\newcommand{\dd}[2][]{\frac{\mathrm{d} #1}{\mathrm{d} #2}}
\newcommand{\pdd}[2][]{\frac{\partial #1}{\partial #2}}
\newcommand{\atanh}{\operatorname{atanh}}
\newcommand{\sech}{\operatorname{sech}}
\newcommand{\keld}{\tilde{\gamma}}
\renewcommand{\Re}{\operatorname{Re}}
\renewcommand{\Im}{\operatorname{Im}}
\newcommand\numberthis{\addtocounter{equation}{1}\tag{\theequation}}

\newcommand{\ket}[1]{\left|#1\right>}
\newcommand{\bra}[1]{\left<#1\right|}
\newcommand{\braket}[2]{\left<#1|#2\right>}
\newcommand{\nn}{\nonumber\\}
\newcommand{\ul}{\underline}
\newcommand{\f}[1]{\mbox{\boldmath$#1$}}
\newcommand{\fk}[1]{\mbox{\boldmath$\scriptstyle#1$}}
\newcommand{\vau}{\mbox{\boldmath$v$}}
\newcommand{\na}{\mbox{\boldmath$\nabla$}}
\newcommand{\bea}{\begin{eqnarray}}
\newcommand{\ea}{\end{eqnarray}}
\newcommand{\eea}{\end{eqnarray}}
\newcommand{\ord}{\,{\cal O}}
\newcommand{\li}{\,\widehat{\cal L}}
\newcommand{\vc}[1]{\mathbf{#1}}
\newcommand{\sumint}[1]
{\begin{array}{c} \\
{{\textstyle\sum}\hspace{-1.1em}{\displaystyle\int}}\\
{\scriptstyle{#1}}
\end{array}}

\title{On the observability of field-assisted birefringent Delbr\"uck scattering}

\author{N.~Ahmadiniaz} 
\affiliation{Helmholtz-Zentrum Dresden-Rossendorf, Bautzner Landstra\ss e 400, 01328 Dresden, Germany} 

\author{M.~Bussmann}
\affiliation{Helmholtz-Zentrum Dresden-Rossendorf, Bautzner Landstra\ss e 400, 01328 Dresden, Germany} 
\affiliation{Center for Advanced Systems Understanding (CASUS), G\"orlitz, Germany}

\author{T. E.~Cowan}
\affiliation{Helmholtz-Zentrum Dresden-Rossendorf, Bautzner Landstra\ss e 400, 01328 Dresden, Germany} 
\affiliation{Institut f\"ur Kern-und Teilchenphysik, Technische Universit\"at Dresden, 01062 Dresden, Germany}

\author{A.~Debus}
\affiliation{Helmholtz-Zentrum Dresden-Rossendorf, Bautzner Landstra\ss e 400, 01328 Dresden, Germany} 

\author{T.~Kluge}
\affiliation{Helmholtz-Zentrum Dresden-Rossendorf, Bautzner Landstra\ss e 400, 01328 Dresden, Germany} 

\author{R.~Sch\"utzhold}
\affiliation{Helmholtz-Zentrum Dresden-Rossendorf, Bautzner Landstra\ss e 400, 01328 Dresden, Germany} 
\affiliation{Institut f\"ur Theoretische Physik, Technische Universit\"at Dresden, 01062 Dresden, Germany}

\begin{abstract}
We consider the scattering of an x-ray free-electron laser (XFEL) beam on the superposition of 
a strong magnetic field $\bf{B}_{\rm ext}$ with the Coulomb field $\bf{E}_{\rm ext}$ 
of a nucleus with charge number $Z$. 
In contrast to pure Delbr\"uck scattering (Coulomb field only), the magnetic field $\bf{B}_{\rm ext}$ 
introduces an asymmetry (i.e., polarization dependence) and renders the effective interaction volume quite 
large, while the nuclear Coulomb field facilitates a significant momentum transfer $\Delta\bf k$. 
For a field strength of $B_{\rm ext}=10^6~\rm T$ (corresponding to an intensity of order $10^{22}~\rm W/cm^2$)
and an XFEL frequency of 24~keV, we find a differential cross section 
$d\sigma/d\Omega\sim10^{-25}~Z^2/(\Delta{\bf k})^2$ in forward direction for one nucleus. 
Thus, this effect might be observable in the near future at facilities such as the 
Helmholtz International Beamline for Extreme Fields (HIBEF) at the European XFEL.  
\end{abstract}

\date{\today} 

\maketitle

%%%%%%%%%%%%%%%%%%%%%%%%%%%%%%%%%%%%%%%%%%%%%%%%%%%%%%%%%%%%%%%%%%%%%%%%%%%%%%%%%%%%%%%%%%%%%%%%%%%%%%%%%%%%%%%%%%
\paragraph{Introduction}
%%%%%%%%%%%%%%%%%%%%%%%%%%%%%%%%%%%%%%%%%%%%%%%%%%%%%%%%%%%%%%%%%%%%%%%%%%%%%%%%%%%%%%%%%%%%%%%%%%%%%%%%%%%%%%%%%%

According to classical electrodynamics, electromagnetic waves in vacuum obey the superposition principle and 
thus do not influence each other. 
Quantum electrodynamics (QED), on the other hand, predicts that they do interact via their coupling to the 
fermionic degrees of freedom \cite{euler-35,euler+heisenberg,karplus+neuman}.  
The difficulties of observing this interaction can be understood by recalling the characteristic scales of QED.
First, the electron mass $m\approx0.51~{\rm MeV}/c^2$ sets an energy scale where the associated 
length scale $\lambdabar=\hbar/(mc)\approx386~\rm fm$ is the %(reduced) 
Compton length.  
Second, the elementary charge $q$ can be used to construct the Schwinger critical field 
\cite{sauter-31,schwinger-51} 
\bea
\label{Schwinger} 
E_{\rm crit}=\frac{m^2c^3}{\hbar q}\approx 1.3\times10^{18}\,\frac{\rm V}{\rm m}
\,,
\ea
where the corresponding magnetic field $B_{\rm crit}=E_{\rm crit}/c$ is given by 
$B_{\rm crit}\approx4.4\times10^9~\rm T$.
It is extremely hard to reach such field strengths in the laboratory, but even stronger 
fields exist in extra-terrestrial environments, 
cf.~\cite{neutron-star,neutron-star-comment,mesz,ruffini}.  

As an exception, the Coulomb field of a nucleus with charge $Zq$ exceeds the field
strength~\eqref{Schwinger} very close to the nucleus, i.e., on a distance of order 
$\ord(\sqrt{Z}\lambdabar)$. 
Such a high field strength helps to observe the interaction of electromagnetic fields
and the scattering of photons at the nuclear Coulomb field (usually referred to as 
Delbr\"uck scattering \cite{meitner-33,bethe+rohrlich,costantini-71}) 
has been observed in several experiments, see, e.g., 
\cite{cheng1,cheng2,milshtein-1,milshtein-2,schumacher-75,falkenberg-92,
DESY-exp,jackson-69,akhm-98,moreh,kahane,rev1,rev2,rev3,rev4,rull-81}.  
In order to probe the high field in the close vicinity of the nucleus, 
these photons had an energy of the order of the electron mass or above. 
As another example, the interaction of the Coulomb fields of two nuclei almost 
colliding with each other at ultra-high energies and the resulting emission of 
a pair of photons has also been observed \cite{atlas}.

%%%%%%%%%%%%%%%%%%%%%%%%%%%%%%%%%%%%%%%%%%%%%%%%%%%%%%%%%%%%%%%%%%%%%%%%%%%%%%%%%%%%%%%%%%%%%%%%%%%%%%%%%%%%%%%%%%
\begin{figure}
\includegraphics[width=0.45\textwidth]{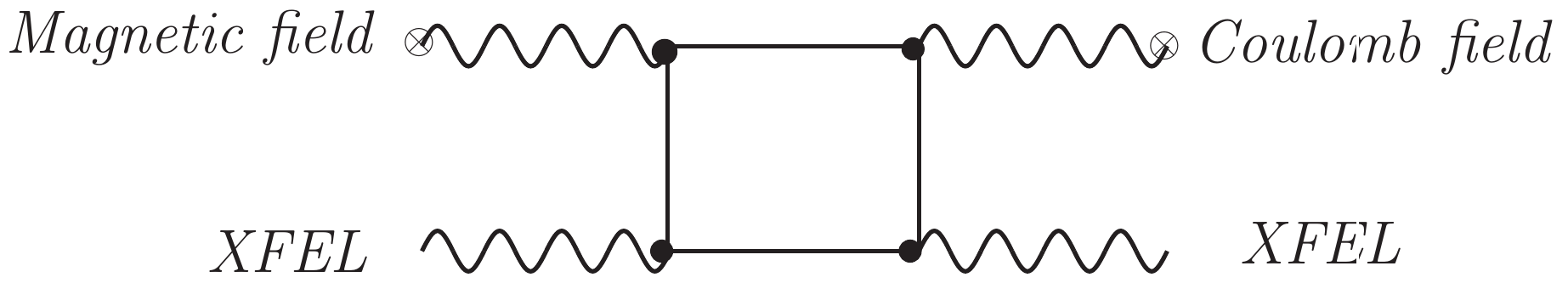}
\caption{Exemplary Feynman diagram of the considered process
(wavy/straight lines represent photons/electrons).}
\label{figure}
\end{figure}
%%%%%%%%%%%%%%%%%%%%%%%%%%%%%%%%%%%%%%%%%%%%%%%%%%%%%%%%%%%%%%%%%%%%%%%%%%%%%%%%%%%%%%%%%%%%%%%%%%%%%%%%%%%%%%%%%%

In contrast to the cases mentioned above, the interaction of electromagnetic 
fields (as predicted by QED) in a regime well below the QED scales $mc^2$, 
$E_{\rm crit}$ and $B_{\rm crit}$ has not been observed yet. 
Prominent proposals for ongoing and planned experiments include the interaction 
of an optical (or near-optical) laser with a (quasi) static magnetic field of 
a few Tesla, see, e.g., \cite{della-16,pvlas,zava-12,zava-06,zava-08,OVAL,BMV,hartman-17,
battesti-18}, 
the interaction of x-ray free electron laser (XFEL) beams among each other or 
with optical lasers, see, e.g., 
\cite{heinzl-06,inada-14,yamaji-16,Schlenvoigt-2016,Karbstein-2016,inada-2017},  
or the interaction of several optical lasers, see, e.g.,  
\cite{dipiazza-07,tommasini-09,tommasini-10,koga-17,
king-12,gies1,gies2,gies3,gies-09,grote-15,ahm-20}. 

Here, we consider a mixed set-up where an XFEL beam is scattered at the combination of a 
strong magnetic field $\bf{B}_{\rm ext}$ superimposed by the Coulomb field $\bf{E}_{\rm ext}$
of a nucleus (as schematically depicted in Fig.~\ref{figure}), see also \cite{dipiazza-08}. 
This scenario offers several advantages:
The nuclear Coulomb field facilitates a significant momentum transfer $\Delta\bf k$ 
while we still obtain a birefringent signal (where the polarization of the XFEL photons flips). 
The crux of our proposal and the prime difference to the aforementioned scenarios involving 
nuclear fields studied previously is that we 
obtain a very large interaction volume whose length and time scales are set by the momentum 
transfer $\Delta\bf k$ and thus well above the Compton length $\lambdabar$. 
Hence, all involved field strengths are sub-critical, i.e., well below~\eqref{Schwinger}. 

%%%%%%%%%%%%%%%%%%%%%%%%%%%%%%%%%%%%%%%%%%%%%%%%%%%%%%%%%%%%%%%%%%%%%%%%%%%%%%%%%%%%%%%%%%%%%%%%%%%%%%%%%%%%%%%%%%
\paragraph{Euler-Heisenberg Lagrangian}%\label{Euler-Heisenberg}
%%%%%%%%%%%%%%%%%%%%%%%%%%%%%%%%%%%%%%%%%%%%%%%%%%%%%%%%%%%%%%%%%%%%%%%%%%%%%%%%%%%%%%%%%%%%%%%%%%%%%%%%%%%%%%%%%%

Since we are considering slowly varying electric $\mathfrak{E}$ and 
magnetic $\mathfrak{B}$ fields well below $E_{\rm crit}$ and $B_{\rm crit}$, 
we may start with the lowest-order Euler-Heisenberg Lagrangian 
($\hbar=c=\epsilon_0=\mu_0=1$) 
\bea
\label{EHL}
\mathfrak{L}
=
\frac{1}{2}\left(\mathfrak{E}^2-\mathfrak{B}^2\right)+
\xi\left[\left(\mathfrak{E}^2-\mathfrak{B}^2\right)^2
+7\left(\mathfrak{E}\cdot\mathfrak{B}\right)^2
\right],
\ea
with a non-linearity set by the parameter 
\bea
\xi
=
\frac{q^4}{360\pi^2 m^4}
=
\frac{2\alpha_{\rm QED}^2}{45m^4}
=
\frac{\alpha_{\rm QED}}{90\pi E_{\rm crit}^2}
\,,
\ea
where $\alpha_{\rm QED}\approx1/137$ is the fine-structure constant
\cite{euler+heisenberg}, see also \cite{review-karbstein-2020,review-king+heinzl-2016,
review-karbstein-2016,review-dipiazza-12,review-dunne-12,akhm-02,Bialynicki-Birula-1970,Toll-PhD,adler70,adler71,adler96}. 
This is a great advantage because we do not have to construct the electron 
propagator whose explicit form is known in special cases only \cite{foot1}, 
e.g., in an electromagnetic plane-wave background which facilitates Volkov solutions 
\cite{volkov,red-65}, 
see also \cite{milstein-05,diphatkei-08,dipiazza-12,dipiazza-14}. 

In the weak-field limit considered here, we neglect quadratic terms $\ord(\xi^2)$. 
Furthermore, we assume that the magnetic field $\bf{B}_{\rm ext}$ is (approximately) constant. 
In addition to this magnetic field $\bf{B}_{\rm ext}$ and the static Coulomb field $\bf{E}_{\rm ext}$ 
of the nucleus, we have the space-time dependent XFEL fields ${\bf E}$ and ${\bf B}$. 
Inserting this split 
$\mathfrak{E}={\bf E}_{\rm ext}+{\bf E}$ and $\mathfrak{B}={\bf B}_{\rm ext}+{\bf B}$
into~\eqref{EHL}, we obtain the effective Lagrangian for the XFEL fields (cf.~\cite{ahm-20}) 
\bea
\label{probe}
\mathfrak{L}_{\rm XFEL}
&=&
\half\left[
{\bf E}\cdot(\mathbb{1}+\delta\epsilon)\cdot{\bf E}-
{\bf B}\cdot(\mathbb{1}-\delta\mu)\cdot{\bf B}
\right] 
\nn
&&
+{\bf E}\cdot\delta\Psi\cdot{\bf B}
\,, 
\ea
with the symmetric permittivity/permeability tensors 
$\delta\epsilon^{ij}=
8\xi E_{\rm ext}^i E_{\rm ext}^j 
+14\xi B_{\rm ext}^i B_{\rm ext}^j 
+4\xi\delta^{ij}({\bf E}_{\rm ext}^2-{\bf B}_{\rm ext}^2)$
and
$\delta\mu^{ij}
=
8\xi B_{\rm ext}^i B_{\rm ext}^j 
+14\xi E_{\rm ext}^i E_{\rm ext}^j 
-4\xi\delta^{ij}({\bf E}_{\rm ext}^2-{\bf B}_{\rm ext}^2)$ 
plus the symmetry-breaking contribution
\bea
\label{symmetry-breaking}
\delta\Psi^{ij}
&=&
-8\xi E_{\rm ext}^i B_{\rm ext}^j 
+14\xi B_{\rm ext}^i E_{\rm ext}^j 
\nn
&&
+14\xi\delta^{ij}({\bf E}_{\rm ext}\cdot{\bf B}_{\rm ext})
\,,
\ea
which describe the polarizability of the QED vacuum. 
Note that the latter tensor is not symmetric 
$\delta\Psi^{ij}\neq\delta\Psi^{ji}$. 

The equations of motion stemming from~\eqref{probe} can be cast into the 
same form as the macroscopic Maxwell equations in a medium 
$\nabla\cdot{\bf D}=0$, $\nabla\cdot{\bf B}=0$, 
$\nabla\times{\bf E}=-\partial_t{\bf B}$, and 
$\nabla\times{\bf H}=\partial_t{\bf D}$, provided that we introduce the 
electric 
${\bf D}=(\mathbb{1}+\delta\epsilon)\cdot{\bf E}+\delta\Psi\cdot{\bf B}$ 
and magnetic displacement fields 
${\bf H}=(\mathbb{1}-\delta\mu)\cdot{\bf B}-\delta\Psi^{\rm T}\cdot{\bf E}$. 

%%%%%%%%%%%%%%%%%%%%%%%%%%%%%%%%%%%%%%%%%%%%%%%%%%%%%%%%%%%%%%%%%%%%%%%%%%%%%%%%%%%%%%%%%%%%%%%%%%%%%%%%%%%%%%%%%%
\paragraph{Scattering Theory}%\label{Scattering}
%%%%%%%%%%%%%%%%%%%%%%%%%%%%%%%%%%%%%%%%%%%%%%%%%%%%%%%%%%%%%%%%%%%%%%%%%%%%%%%%%%%%%%%%%%%%%%%%%%%%%%%%%%%%%%%%%%

Now we may calculate the scattering of the XFEL beam with standard approaches, 
see, e.g., \cite{jackson}. 
Combining the above Maxwell equations to 
\bea
\Box{\bf D}
=
\nabla\times\left[\nabla\times({\bf D}-{\bf E})\right]
+\partial_t\left[\nabla\times({\bf H}-{\bf B})\right]
={\bf J}^{\rm eff}\,,
\ea
where the effective source term $ {\bf J}^{\rm eff}$ 
on the right-hand side is small, allows us to employ the 
Born approximation.
To this end, we split the XFEL field ${\bf D}$ into an ingoing plane wave 
${\bf D}^{\rm in}$ plus a small scattering contribution ${\bf D}^{\rm out}$ 
induced by vacuum polarizability $\delta\epsilon$, $\delta\mu$, and $\delta\Psi$.
Assuming a stationary time-dependence $e^{-i\omega t}$ for the XFEL field 
($\delta\epsilon$, $\delta\mu$, and $\delta\Psi$ are static), we find 
\bea
\Box{\bf D}_\omega^{\rm out}
&=&
-\left(\nabla^2+\omega^2\right){\bf D}_\omega^{\rm out}
=
{\bf J}_\omega^{\rm eff}
\nn
&=&
\nabla\times\left[\nabla\times(
\delta\epsilon\cdot{\bf E}_\omega^{\rm in}+\delta\Psi\cdot{\bf B}_\omega^{\rm in}
)\right]
\nn
&&
+
i\omega\nabla\times(
\delta\mu\cdot{\bf B}_\omega^{\rm in}+\delta\Psi^{\rm T}\cdot{\bf E}_\omega^{\rm in}
)\,,
\ea
where we may approximate ${\bf E}_\omega^{\rm in}\approx{\bf D}_\omega^{\rm in}$
and ${\bf B}_\omega^{\rm in}\approx{\bf H}_\omega^{\rm in}$
on the right-hand side within the Born approximation.
As usual, solving this Helmholtz differential equation for ${\bf D}_\omega^{\rm out}$
with the standard Greens function and considering the behavior at large spatial 
distances, we find the differential cross section $d\sigma/d\Omega=|{\mathfrak A}|^2$
with the scattering amplitude \cite{jackson}
\bea
{\mathfrak A}=\frac{1}{4\pi|{\bf D}_\omega^{\rm in}|}\,{\bf e}_{\rm out}\cdot
\int d^3r\,\exp\{-i{\bf k}_{\rm out}\cdot{\bf r}\}\,{\bf J}_\omega^{\rm eff}
\,,
\ea
where ${\bf k}_{\rm out}$ is the wave-number and ${\bf e}_{\rm out}$ 
the polarization unit vector of the scattered (outgoing) XFEL radiation.  

%%%%%%%%%%%%%%%%%%%%%%%%%%%%%%%%%%%%%%%%%%%%%%%%%%%%%%%%%%%%%%%%%%%%%%%%%%%%%%%%%%%%%%%%%%%%%%%%%%%%%%%%%%%%%%%%%%
\paragraph{Field-Assisted Scattering}%\label{Assisted}
%%%%%%%%%%%%%%%%%%%%%%%%%%%%%%%%%%%%%%%%%%%%%%%%%%%%%%%%%%%%%%%%%%%%%%%%%%%%%%%%%%%%%%%%%%%%%%%%%%%%%%%%%%%%%%%%%%

The effective source term ${\bf J}_\omega^{\rm eff}$ contains 
contributions from $\delta\epsilon$, $\delta\mu$, and $\delta\Psi$ 
which add up to give the full amplitude
\bea
\label{total-amplitude} 
{\mathfrak A}
&=&
\frac{\omega^2}{4\pi}\int d^3r\,\exp\{i\Delta{\bf k}\cdot{\bf r}\}
\big[
{\bf e}_{\rm out}\cdot\delta\epsilon\cdot{\bf e}_{\rm in}
\nn
&&
+{\bf e}_{\rm out}\cdot\delta\Psi\cdot({\bf n}_{\rm in}\times{\bf e}_{\rm in}) 
+{\bf e}_{\rm in}\cdot\delta\Psi\cdot({\bf n}_{\rm out}\times{\bf e}_{\rm out}) 
\nn
&&+
({\bf n}_{\rm out}\times{\bf e}_{\rm out})\cdot\delta\mu\cdot({\bf n}_{\rm in}\times{\bf e}_{\rm in}) 
\big]
\,.
\ea
Here ${\bf n}_{\rm in}$ and ${\bf n}_{\rm out}$ are the initial and final 
propagation directions (${\bf k}_{\rm in}=\omega{\bf n}_{\rm in}$ and 
${\bf k}_{\rm out}=\omega{\bf n}_{\rm out}$), respectively, while 
${\bf e}_{\rm in}$ and ${\bf e}_{\rm out}$ denote their polarizations. 
As usual in scattering theory, the oscillating phase is governed by the 
momentum transfer $\Delta{\bf k}={\bf k}_{\rm in}-{\bf k}_{\rm out}$. 

In the following, we only consider the terms stemming from the combined impact 
of the magnetic field ${\bf B}_{\rm ext}$ and the nuclear Coulomb field 
${\bf E}_{\rm ext}$. 
Keeping only those cross terms, just the symmetry-breaking contribution 
$\delta\Psi$ in~\eqref{symmetry-breaking} survives, i.e., we focus on the 
second line of Eq.~\eqref{total-amplitude}. 

Since ${\bf B}_{\rm ext}$ is (nearly) constant, the spatial integral 
yields the Fourier transform of the nuclear Coulomb field 
${\bf E}_{\rm ext}({\bf r})={\bf e}_r Q/(4\pi{\bf r}^2)$ with the charge $Q=Zq$
\bea
\label{Fourier-Coulomb}
\int d^3r\,e^{i\Delta{\bf k}\cdot{\bf r}}\,\frac{Q}{4\pi{\bf r}^2}\,{\bf e}_r
= 
iQ\,\frac{\Delta{\bf k}}{(\Delta{\bf k})^2}
\,.
\ea
An important point here is that the $1/r^2$ scaling from the Coulomb field 
cancels the $r^2$ volume factor in the $d^3r$ integration. 
As a consequence, the spatial integration is cut off by the momentum transfer 
$\Delta{\bf k}$ resulting in a very large interaction volume -- which may even 
span many XFEL wavelengths for small scattering angles 
$|{\bf n}_{\rm in}-{\bf n}_{\rm out}|\ll1$, i.e., in forward direction. 
Of course, at some point the approximation of a constant ${\bf B}_{\rm ext}$ 
breaks down. 

%%%%%%%%%%%%%%%%%%%%%%%%%%%%%%%%%%%%%%%%%%%%%%%%%%%%%%%%%%%%%%%%%%%%%%%%%%%%%%%%%%%%%%%%%%%%%%%%%%%%%%%%%%%%%%%%%%
\paragraph{Forward Scattering}%\label{Assisted}
%%%%%%%%%%%%%%%%%%%%%%%%%%%%%%%%%%%%%%%%%%%%%%%%%%%%%%%%%%%%%%%%%%%%%%%%%%%%%%%%%%%%%%%%%%%%%%%%%%%%%%%%%%%%%%%%%%

The large interaction volume mentioned above goes along with a peak in forward direction, i.e., 
for small $\Delta{\bf k}$, where the signal is enhanced as $\sim1/|\Delta{\bf k}|$.
Since the pre-factor is quite small, let us focus on this leading-order contribution 
$\sim1/|\Delta{\bf k}|$. 
Thus, we approximate ${\bf n}_{\rm in}\approx{\bf n}_{\rm out}\to{\bf n}$ in order to 
simplify the expressions. 
In this limit, the isotropic contribution 
$14\xi\delta^{ij}({\bf E}_{\rm ext}\cdot{\bf B}_{\rm ext})$
in~\eqref{symmetry-breaking} cancels and we are left with the anisotropic terms. 
For the birefringent signal, where ${\bf e}_{\rm in}$ and ${\bf e}_{\rm out}$ 
are (nearly) orthogonal, we may approximate 
${\bf e}_{\rm out}\approx\pm{\bf n}\times{\bf e}_{\rm in}$ and 
${\bf e}_{\rm in}\approx\mp{\bf n}\times{\bf e}_{\rm out}$ which 
simplifies the integrand in~\eqref{total-amplitude} to 
$\pm[{\bf e}_{\rm out}\cdot\delta\Psi\cdot{\bf e}_{\rm out}
-{\bf e}_{\rm in}\cdot\delta\Psi\cdot{\bf e}_{\rm in}]$. 
Altogether, we get the birefringent amplitude 
\bea
\label{perp}
{\mathfrak A}_{\delta\Psi}^\perp 
&=&
\pm6i\xi\,\frac{Q}{(\Delta{\bf k})^2}\,
\frac{\omega^2}{4\pi}
\left[
({\bf e}_{\rm out}\cdot{\bf B}_{\rm ext})
({\bf e}_{\rm out}\cdot\Delta{\bf k}) 
\right.
\nn
&&
\left.
-
({\bf e}_{\rm in}\cdot{\bf B}_{\rm ext})
({\bf e}_{\rm in}\cdot\Delta{\bf k}) 
\right]
\,.
\ea
Thus, one way to obtain a maximum birefringent signal would be to align 
the momentum transfer $\Delta{\bf k}$ with ${\bf B}_{\rm ext}$ as well as 
either ${\bf e}_{\rm in}$ or ${\bf e}_{\rm out}$, for example. 

In this case, we find the amplitude (up to a sign)
\bea
\label{perp-signal} 
{\mathfrak A}_{\delta\Psi}^\perp 
=
6i\xi B_{\rm ext}\,\frac{Q}{|\Delta{\bf k}|}\,
\frac{\omega^2}{4\pi}
=
i\,\frac{\alpha_{\rm QED}^2}{15\pi}\,
\frac{qB_{\rm ext}}{m^2}\,
\frac{\omega^2}{m^2}\,
\frac{Z}{|\Delta{\bf k}|}
\,.
\ea
Apart from a numerical pre-factor, this amplitude scales with the ratios of the 
optical laser field strength $B_{\rm ext}$ over the critical field strength 
$B_{\rm crit}\approx4.4\times10^9~\rm T$ and the square of the XFEL frequency $\omega$ in comparison to the electron mass $m$.
%
%XFEL frequency $\omega$ compared to electron mass $m$ squared.
%

%%%%%%%%%%%%%%%%%%%%%%%%%%%%%%%%%%%%%%%%%%%%%%%%%%%%%%%%%%%%%%%%%%%%%%%%%%%%%%%%%%%%%%%%%%%%%%%%%%%%%%%%%%%%%%%%%%
\paragraph{Detectability} 
%%%%%%%%%%%%%%%%%%%%%%%%%%%%%%%%%%%%%%%%%%%%%%%%%%%%%%%%%%%%%%%%%%%%%%%%%%%%%%%%%%%%%%%%%%%%%%%%%%%%%%%%%%%%%%%%%%

Since the amplitude~\eqref{perp-signal} for field-assisted scattering is proportional to $B_{\rm ext}$ 
and $\omega^2$,  it is desirable to have a strong magnetic field and a large XFEL frequency.
Ultra-high field strengths $B_{\rm ext}=10^6~\rm T$ can be reached in the focus of an ultra-strong optical 
or near optical laser with an intensity of order $10^{22}~\rm W/cm^2$ in a colliding-beam set-up.  
Then, inserting an XFEL frequency of $\omega=24~\rm keV$, which is still within 
the range of the European XFEL \cite{xfel-data,Schlenvoigt-2016,xfel-URL}, we obtain an amplitude of around 
${\mathfrak A}_{\delta\Psi}^\perp\approx5\times10^{-13}Z/|\Delta{\bf k}|$ 
which yields the birefringent differential cross section in forward direction 
\bea
\label{cross-section-perp} 
\frac{d\sigma_{\delta\Psi}^\perp}{d\Omega}=\left|{\mathfrak A}_{\delta\Psi}^\perp\right|^2
\sim10^{-25}\,\frac{Z^2}{(\Delta{\bf k})^2}
\,.
\ea
For a lower frequency of $\omega=1~\rm keV$, the amplitude would be reduced to 
${\mathfrak A}_{\delta\Psi}^\perp\approx10^{-15}Z/|\Delta{\bf k}|$ corresponding to a cross section 
of $10^{-30}Z^2/(\Delta{\bf k})^2$. 

The suppression in~\eqref{cross-section-perp} by more than twenty orders of magnitude is roughly 
comparable to other proposals for vacuum birefringence experiments, see, e.g., 
\cite{heinzl-06,Schlenvoigt-2016}. 
Yet, this suppression does not imply that the effect is beyond reach. 
In order to demonstrate that, let us discuss two main enhancement factors.
The first enhancement factors is a large number $\ord(10^{11})$ of polarized XFEL photons, 
analogous to other vacuum birefringence proposals, see, e.g., \cite{heinzl-06,Schlenvoigt-2016}. 
In addition, a large number $N$ of nuclei represents another enhancement factor of our set-up. 
Note that the peak $\propto1/(\Delta{\bf k})^2$ in forward direction, after integrating over the solid 
angle $d\Omega$, does only yield a weak (i.e., logarithmic) enhancement. 
In addition, as explained after \eqref{Fourier-Coulomb}, this $1/(\Delta{\bf k})^2$ 
behavior is only valid up to minimum values of $|\Delta{\bf k}|$ set by the size of the laser focus.
For even smaller $|\Delta{\bf k}|$, one would have to include the Fourier transform of the dependence of 
${\bf B}_{\rm ext}({\bf r})$ or ${\bf B}_{\rm ext}(t,{\bf r})$ as an effective form factor.

%%%%%%%%%%%%%%%%%%%%%%%%%%%%%%%%%%%%%%%%%%%%%%%%%%%%%%%%%%%%%%%%%%%%%%%%%%%%%%%%%%%%%%%%%%%%%%%%%%%%%%%%%%%%%%%%%%
\begin{figure}
\includegraphics[width=0.41\textwidth]{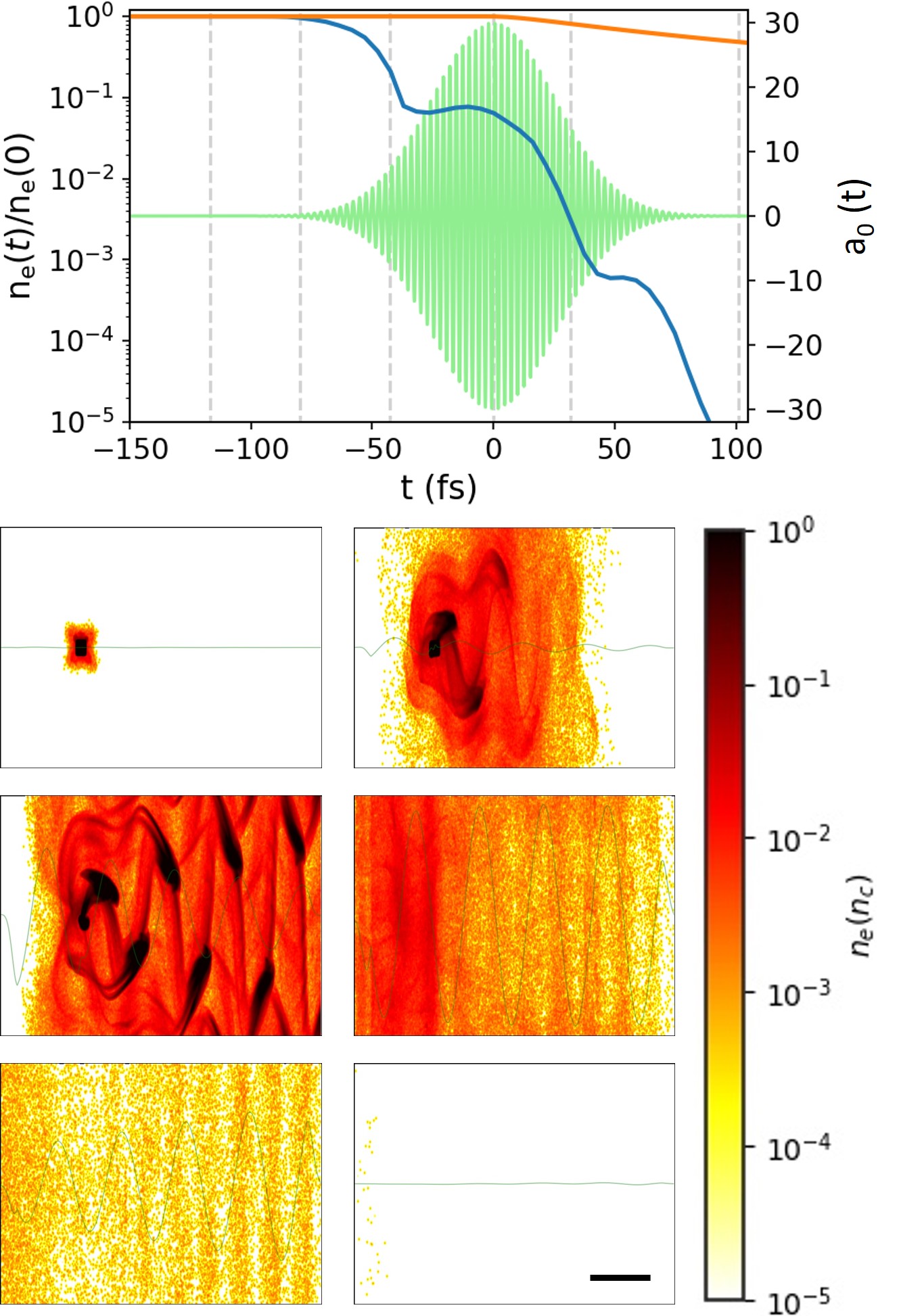}
\caption{
In an exemplary 3D particle-in-cell (PIC) simulation ({\em PIConGPU}), 
a cubic carbon cluster target with 100~nm side length 
and an (initial) electron density $n_e$ of $5\times10^{23}~\rm cm^{-3}$ (i.e., $n_e=290~n_c$)
%$n_e=290n_c~(5\times10^{23}~\rm cm^{-3})$ 
was irradiated by a short pulse laser with 800~nm wavelength, 
30~fs width, $1.946\times10^{21}~\rm W/cm^2$ intensity (i.e., $a_0=30$), incident from the left. 
In order to assess the number of electrons and ions a transversely oriented XFEL probe beam would observe, 
we integrated their respective densities perpendicular to the plane shown in the lower panels in a 
200~nm~$\times$~200~nm square area. 
The resulting normalized temporal evolution is plotted in the top: 
electrons blue, ions orange and the laser electric field is shown in green for reference. 
The vertical gray dashed lines indicate the times for which the lower panels show the electron density and 
laser electric field. 
From top left to bottom right these are $-117$~fs, $-80$~fs, $-43$~fs, $+32$~fs, and $+101$~fs. 
We see that the electron number in the interaction volume drops by five orders of magnitude while 
the ion number only decreases moderately (due to the Coulomb explosion). 
The simulation box has periodic boundaries in the dimensions transverse to the laser; 
the black scale bar in the bottom right is 500~nm long.}
\label{fig-pic}
\end{figure}
%%%%%%%%%%%%%%%%%%%%%%%%%%%%%%%%%%%%%%%%%%%%%%%%%%%%%%%%%%%%%%%%%%%%%%%%%%%%%%%%%%%%%%%%%%%%%%%%%%%%%%%%%%%%%%%%%%

As one possible scenario (see Fig.~\ref{fig-pic}), one may consider a cubic cluster with 
an edge length of 100~nm with typical solid-state density made of carbon, for example. 
Then, applying a pre-pulse with a high intensity a bit below $\ord(10^{22}~\rm W/cm^2)$, 
one may blow out almost all electrons, leaving behind $N=\ord(10^8)$ ionized nuclei.  
Shortly afterwards, before the Coulomb explosion of the remaining nuclei, one would have them interact with 
the XFEL superimposed by the main pulse (in the form of a colliding beam set-up, for example).
To exploit the peak in forward direction, let us assume a small momentum transfer $\Delta\bf{k}$
in the eV regime (corresponding to scattering angles $\vartheta$ of order millirad).  
In this case, the amplitudes from the $N=\ord(10^8)$ nuclei would have basically the same phase 
and thus add up coherently, effectively acting as one giant nucleus with charge $Z_{\rm eff}=NZ$. 
For much larger $\Delta\bf{k}$, one would have to include the spatial distribution of the nuclei 
in analogy to \eqref{Fourier-Coulomb}.

For smaller clusters (e.g., 50~nm), the electrons would already be blown out before the laser 
pulse reaches its peak (see the Appendix).
Thus, in this second scenario, one could avoid the ``pre-pulse -- main pulse'' sequence and use 
the same laser pulse for ionization and assisting Delbr\"uck scattering.
However, in this case one should also take into account the electric field component of the optical laser. 
This electric field of the laser would then generate additional contributions in $\delta\epsilon$ and $\delta\mu$ 
after combining it with the nuclear Coulomb field. 
Inserting those additional contributions in $\delta\epsilon$ and $\delta\mu$ into the full 
amplitude~\eqref{total-amplitude}, we find that they exactly cancel the terms from $\delta\Psi$ 
if the optical laser and the XFEL propagate in the {\em same} direction. 
This cancellation does also occur in the case without the nuclear Coulomb field and demonstrates an important 
difference between a propagating plane wave (crossed fields) and a pure magnetic field. 
To avoid this cancellation and obtain a birefringent signal, the XFEL should propagate at a finite angle 
(e.g., perpendicular) to the propagation direction as well as the magnetic field component of the optical laser. 

As a third scenario, one could envisage an optical laser focus (expelling the electrons) co-propagating
with the XFEL pulse through a less dense medium, in analogy to laser wake-field acceleration \cite{Esarey2009}.
In the usual set-up, the optical laser would co-propagate with the created fully blown-out plasma cavity 
and thus also with the XFEL.
However, this would again lead to the cancellation problem discussed above.
To overcome this problem, one could use an optical laser with a propagation direction different from that of
the XFEL, whose {\em focus} is co-propagating with the XFEL pulse.
This can be achieved by laser pulse-front shaping techniques \cite{Debus2019,Steiniger2019,Steiniger2014,Debus2010}.
Inserting typical numbers such as a density of order $10^{19}~\rm cm^{-3}$, the number $N$ of ions within the
interaction region, i.e., the laser focus with a 2.5~$\mu$m spot size, travelling over a distance of order 
millimeter, would be about two to three orders of magnitude larger $N=\ord(6\times10^{10})$ compared to the 
first
 scenario.
However, presumably not all of these ions would contribute coherently to the scattering amplitude in view of 
the larger spatial extent (again compared to the first and second scenario). 
As an advantage, some of the background processes (see the Appendix) due to the residual electrons and the 
residual radiation are minimized in the third scenario. 
In summary,  the three scenarios offer different advantages and drawbacks, which should be compared carefully 
for designing an experimental realization.

Finally, the measurement of the birefringent signal, i.e., the detection of the XFEL photons with flipped 
polarization could be achieved in complete analogy to other vacuum birefringence proposals, see, e.g., 
\cite{heinzl-06,Schlenvoigt-2016}. 

%%%%%%%%%%%%%%%%%%%%%%%%%%%%%%%%%%%%%%%%%%%%%%%%%%%%%%%%%%%%%%%%%%%%%%%%%%%%%%%%%%%%%%%%%%%%%%%%%%%%%%%%%%%%%%%%%%
\paragraph{Conclusions}
%%%%%%%%%%%%%%%%%%%%%%%%%%%%%%%%%%%%%%%%%%%%%%%%%%%%%%%%%%%%%%%%%%%%%%%%%%%%%%%%%%%%%%%%%%%%%%%%%%%%%%%%%%%%%%%%%%

As an example for the QED vacuum nonlinearity, we calculate the scattering of XFEL photons at the combined field 
of a nucleus plus an external magnetic field (e.g., generated by an optical laser focus), see Fig.~\ref{figure}. 
In contrast to previous work involving nuclear fields, such as pure Delbr\"uck scattering, this scenario yields 
a large interaction volume, which goes along with a peak of the differential cross section in forward direction 
$d\sigma/d\Omega\sim1/(\Delta{\bf k})^2$. 

As another distinction, the scales relevant to our scenario are well below the characteristic scales of QED
mentioned in the Introduction, i.e., the critical field strength~\eqref{Schwinger} and the electron mass $m$. 
Thus, our scenario is more within the regime where the approximation of a classical electromagnetic field 
applies -- especially for the coherent superposition of the signal from many nuclei -- 
than within the particle (photon) picture often associated with pure Delbr\"uck scattering. 

In addition to the normal polarization conserving scattering (see the Appendix), 
we obtain a birefringent signal~\eqref{perp-signal} whose amplitude ${\mathfrak A}_{\delta\Psi}^\perp$ 
is just a little bit smaller. 
Besides the anisotropy induced by the magnetic field and the aforementioned peak in forward direction, 
this birefringence may be used to distinguish the process of field-assisted Delbr\"uck scattering considered 
here from other background processes discussed in the Appendix.  
Using a large number of XFEL photons \cite{xfel-URL,xfel-data} and a large number of nuclei, 
we show that it might be possible to overcome the suppression of the signal~\eqref{perp-signal} by more than 
twenty orders of magnitude such that this effect might be observable in the near future at facilities such as 
the Helmholtz International Beamline for Extreme Fields (HIBEF) at the European XFEL \cite{hibef-URL}.  

%%%%%%%%%%%%%%%%%%%%%%%%%%%%%%%%%%%%%%%%%%%%%%%%%%%%%%%%%%%%%%%%%%%%%%%%%%%%%%%%%%%%%%%%%%%%%%%%%%%%%%%%%%%%%%%%%%
\acknowledgments 
%%%%%%%%%%%%%%%%%%%%%%%%%%%%%%%%%%%%%%%%%%%%%%%%%%%%%%%%%%%%%%%%%%%%%%%%%%%%%%%%%%%%%%%%%%%%%%%%%%%%%%%%%%%%%%%%%%

We would like to thank C.~Schubert as well as R.~Sauerbrey, R.~Shaisultanov, G.~Torgrimsson, 
and other colleagues from the HZDR for helpful discussions. 
This work was partially funded by 
the Deutsche Forschungsgemeinschaft (DFG, German Research Foundation) -- Project-ID 278162697 -- SFB 1242; 
as well as by 
the Center of Advanced Systems Understanding (CASUS) which is financed by 
Germany’s Federal Ministry of Education and Research (BMBF) and by the 
Saxon Ministry for Science, Culture and Tourism (SMWK) 
with tax funds on the basis of the budget approved by the Saxon State Parliament.

%\appendix 

%%%%%%%%%%%%%%%%%%%%%%%%%%%%%%%%%%%%%%%%%%%%%%%%%%%%%%%%%%%%%%%%%%%%%%%%%%%%%%%%%%%%%%%%%%%%%%%%%%%%%%%%%%%%%%%%%%
\section{Appendix: Backward Scattering}%\label{Assisted}
%%%%%%%%%%%%%%%%%%%%%%%%%%%%%%%%%%%%%%%%%%%%%%%%%%%%%%%%%%%%%%%%%%%%%%%%%%%%%%%%%%%%%%%%%%%%%%%%%%%%%%%%%%%%%%%%%%

In contrast to the case of forward scattering discussed above, let us study the opposite case of backward 
scattering ${\bf n}_{\rm in}=-{\bf n}_{\rm out}$ for comparison. 
Again focusing on the birefringent signal ${\bf e}_{\rm in}\perp{\bf e}_{\rm out}$, we find 
\bea
\label{perp-signal-back} 
{\mathfrak A}_{\delta\Psi}^\perp 
=
i\,\frac{7\alpha_{\rm QED}^2}{45\pi}\,
\frac{q{\bf B}_{\rm ext}\cdot{\bf n}_{\rm out}}{m^2}\,
\frac{Z\omega}{m^2}\,
\,.
\ea
In contrast to forward scattering, here the isotropic part 
$14\xi\delta^{ij}({\bf E}_{\rm ext}\cdot{\bf B}_{\rm ext})$ in~\eqref{symmetry-breaking} 
contributes while the anisotropic terms cancel. 
As a consequence, the maximum signal is obtained if the magnetic field ${\bf B}_{\rm ext}$
is aligned with the XFEL propagation direction ${\bf n}_{\rm in}$ and thus perpendicular to 
both ${\bf e}_{\rm in}$ and ${\bf e}_{\rm out}$. 
%

%%%%%%%%%%%%%%%%%%%%%%%%%%%%%%%%%%%%%%%%%%%%%%%%%%%%%%%%%%%%%%%%%%%%%%%%%%%%%%%%%%%%%%%%%%%%%%%%%%%%%%%%%%%%%%%%%%
\section{Appendix: Parallel Polarizations}
%%%%%%%%%%%%%%%%%%%%%%%%%%%%%%%%%%%%%%%%%%%%%%%%%%%%%%%%%%%%%%%%%%%%%%%%%%%%%%%%%%%%%%%%%%%%%%%%%%%%%%%%%%%%%%%%%%

To complete our analysis, we also consider the case of parallel polarizations, even though this signal is 
probably much harder to distinguish from the background processes (see below). 
Interestingly, for backward scattering ${\bf n}_{\rm in}=-{\bf n}_{\rm out}$, the field-assisted contribution 
from $\delta\Psi$ vanishes. 
In forward direction, we get a $1/|\Delta{\bf k}|$ peak in analogy to the birefringent signal~\eqref{perp}
\bea
\label{parallel} 
{\mathfrak A}_{\delta\Psi}^\|
&=&
i\xi\,\frac{Q}{(\Delta{\bf k})^2}\,
\frac{\omega^2}{4\pi}
\left[
28({\bf e}\cdot{\bf B}_{\rm ext})({\bf n}\times{\bf e})\cdot\Delta{\bf k}
\right.
\nn
&&
\left.
-16({\bf e}\cdot\Delta{\bf k})({\bf n}\times{\bf e})\cdot{\bf B}_{\rm ext}
\right]
\,,
\ea
where we have set ${\bf e}_{\rm in}$ and ${\bf e}_{\rm out}$ to the overall XFEL polarization unit vector 
${\bf e}$. 
For both polarizations (perpendicular and parallel), the leading-order amplitudes~\eqref{perp} and 
\eqref{parallel} in forward direction vanish identically if the XFEL propagation direction $\bf n$ 
is parallel to the magnetic field ${\bf B}_{\rm ext}$. 
In contrast to the birefringent signal~\eqref{perp}, the above amplitude~\eqref{parallel} assumes its 
maximum contribution if the polarization $\bf e$ (i.e., the electric field) of the XFEL is parallel to 
the magnetic field ${\bf B}_{\rm ext}$  
(and thus the XFEL propagation direction $\bf n$ perpendicular to ${\bf B}_{\rm ext}$), 
while the momentum transfer $\Delta{\bf k}$ is perpendicular to both,
i.e., in the ``equatorial plane''. 

%%%%%%%%%%%%%%%%%%%%%%%%%%%%%%%%%%%%%%%%%%%%%%%%%%%%%%%%%%%%%%%%%%%%%%%%%%%%%%%%%%%%%%%%%%%%%%%%%%%%%%%%%%%%%%%%%%
\section{Appendix: Background Processes}%\label{Background}
%%%%%%%%%%%%%%%%%%%%%%%%%%%%%%%%%%%%%%%%%%%%%%%%%%%%%%%%%%%%%%%%%%%%%%%%%%%%%%%%%%%%%%%%%%%%%%%%%%%%%%%%%%%%%%%%%%

Of course, finding such small differential cross sections~\eqref{cross-section-perp}
necessitates the careful estimate of potential background processes.

%%%%%%%%%%%%%%%%%%%%%%%%%%%%%%%%%%%%%%%%%%%%%%%%%%%%%%%%%%%%%%%%%%%%%%%%%%%%%%%%%%%%%%%%%%%%%%%%%%%%%%%%%%%%%%%%%%
\subsection{Nuclear Thomson Scattering}
%%%%%%%%%%%%%%%%%%%%%%%%%%%%%%%%%%%%%%%%%%%%%%%%%%%%%%%%%%%%%%%%%%%%%%%%%%%%%%%%%%%%%%%%%%%%%%%%%%%%%%%%%%%%%%%%%%
%
Perhaps the most obvious candidate is ordinary Thomson scattering of the XFEL photons by the nuclei themselves. 
This process is suppressed for heavy nuclei as its amplitude scales inversely proportional to the mass $M$ 
of the nucleus %${\mathfrak A}_{\rm T}=\alpha_{\rm QED}Z^2/M$. 
\bea
{\mathfrak A}_{\rm T}
=
\alpha_{\rm QED}\,\frac{Z^2}{M} 
\,.
\ea
Inserting a typical mass of 0.9~GeV per nucleon, we find that ${\mathfrak A}_{\rm T}$ 
actually becomes comparable to the parallel polarization signal ${\mathfrak A}_{\delta\Psi}^\|$ 
in~\eqref{parallel} for a momentum transfer in the eV~regime.
However, Thomson scattering ${\mathfrak A}_{\rm T}$ does not display the characteristic 
(anisotropic) dependence on $\Delta{\bf k}$ and ${\bf B}_{\rm ext}$, for example, and does also not 
generate a birefringent signal. 

%%%%%%%%%%%%%%%%%%%%%%%%%%%%%%%%%%%%%%%%%%%%%%%%%%%%%%%%%%%%%%%%%%%%%%%%%%%%%%%%%%%%%%%%%%%%%%%%%%%%%%%%%%%%%%%%%%
\subsection{Nuclear Resonances} 
%%%%%%%%%%%%%%%%%%%%%%%%%%%%%%%%%%%%%%%%%%%%%%%%%%%%%%%%%%%%%%%%%%%%%%%%%%%%%%%%%%%%%%%%%%%%%%%%%%%%%%%%%%%%%%%%%%
%
Similar arguments apply to the scattering of photons by the nuclei via coupling to their internal degrees of 
freedom, usually referred to as nuclear resonances. 
Of course, this process does depend on the concrete nuclear structure, such as the nuclear resonance frequencies, 
which are typically way above the XFEL frequency.
As detailed in \cite{rev4}, for example, the giant dipole resonance (see, e.g., \cite{baldwin-47}) 
dominates in such a case -- but it is still negligible in comparison to ${\mathfrak A}_{\rm T}$ 
for energies well below the MeV regime. 
Furthermore, these nuclear resonances do not generate a birefringent signal to lowest order \cite{rev4}. 

%%%%%%%%%%%%%%%%%%%%%%%%%%%%%%%%%%%%%%%%%%%%%%%%%%%%%%%%%%%%%%%%%%%%%%%%%%%%%%%%%%%%%%%%%%%%%%%%%%%%%%%%%%%%%%%%%%
\subsection{Delbr\"uck Scattering}
%%%%%%%%%%%%%%%%%%%%%%%%%%%%%%%%%%%%%%%%%%%%%%%%%%%%%%%%%%%%%%%%%%%%%%%%%%%%%%%%%%%%%%%%%%%%%%%%%%%%%%%%%%%%%%%%%%
%
Of course, another background process is pure Delbr\"uck scattering,
i.e., the interaction of the XFEL beam with the nuclear Coulomb field only.
If we estimate the amplitude with the approach described above
(in the absence of the optical laser), we find that $\delta\Psi$ vanishes 
while $\delta\epsilon$ and $\delta\mu$ scale as $1/r^4$.
Thus, in contrast to ${\mathfrak A}_{\delta\Psi}$ considered here, 
the effective interaction volume would not be large and the spatial integral 
would be dominated at small (instead if large) radii. 
Formally, this integral would even diverge for $r\to0$, but this is an 
artifact of the used approach based on the leading-order Euler-Heisenberg 
Lagriangian~\eqref{EHL}, which is no longer applicable for very small 
radii since the Coulomb field ${\bf E}_{\rm ext}({\bf r})$ and its gradients 
become large. 
Even though it would be a very interesting and non-trivial problem to 
study Delbr\"uck scattering in this regime, let us obtain a very 
rough estimate \cite{foot2}  
by cutting off the $r$-integral at the Compton length $\lambdabar$
\bea
{\mathfrak A}_{\rm D}
\sim
Z^2
\frac{\alpha_{\rm QED}^3}{m}\,
\frac{\omega^2}{m^2}
\,.
\ea
Thus, pure Delbr\"uck scattering is negligible in comparison to the signal 
${\mathfrak A}_{\delta\Psi}$ considered here. 

%%%%%%%%%%%%%%%%%%%%%%%%%%%%%%%%%%%%%%%%%%%%%%%%%%%%%%%%%%%%%%%%%%%%%%%%%%%%%%%%%%%%%%%%%%%%%%%%%%%%%%%%%%%%%%%%%%
\subsection{Magnetic Field}  
%%%%%%%%%%%%%%%%%%%%%%%%%%%%%%%%%%%%%%%%%%%%%%%%%%%%%%%%%%%%%%%%%%%%%%%%%%%%%%%%%%%%%%%%%%%%%%%%%%%%%%%%%%%%%%%%%%
%
Going to the other limit, the interaction of the XFEL beam with an optical 
laser focus only (i.e., without the nucleus) has already been considered in 
several works, see, e.g., \cite{heinzl-06,tommasini-09,tommasini-10,king-12}.
Since a constant ${\bf B}_{\rm ext}$ would generate constant $\delta\epsilon$ and 
$\delta\mu$ such that the spatial integral would yield $\delta^3(\Delta{\bf k})$,
the possible momentum transfer $\Delta{\bf k}$ is strongly limited by the spatial 
and temporal inhomogeneities of the optical laser focus, see also 
\cite{dipiazza-08,milstein-05,dipiazza-07,dipiazza-12,dipiazza-14}.
As another option for discriminating, this process would be independent of the 
charge $Z$ and the number $N$ of nuclei. 
Note that the optical laser focus can also generate a birefringent signal.  

%%%%%%%%%%%%%%%%%%%%%%%%%%%%%%%%%%%%%%%%%%%%%%%%%%%%%%%%%%%%%%%%%%%%%%%%%%%%%%%%%%%%%%%%%%%%%%%%%%%%%%%%%%%%%%%%%%
\subsection{Electronic Compton Scattering} 
%%%%%%%%%%%%%%%%%%%%%%%%%%%%%%%%%%%%%%%%%%%%%%%%%%%%%%%%%%%%%%%%%%%%%%%%%%%%%%%%%%%%%%%%%%%%%%%%%%%%%%%%%%%%%%%%%%
%
Another important background process is Compton scattering from the residual electrons. 
According to the particle-in-cell (PIC) simulations displayed in Fig.~\ref{fig-pic}, for example, 
the electron number in the interaction volume 
drops by a factor of $10^{-5}$ or more, but there may still be a few $\ord(10^4)$ electrons left.
Thus, even assuming that the amplitudes from the electrons add up coherently, their lowest-order 
Thomson scattering signal would already be below the nuclear Thomson scattering 
${\mathfrak A}_{\rm T}$ as well as the signal ${\mathfrak A}_{\delta\Psi}$ considered here.
However, this Thomson scattering ${\mathfrak A}_{\rm T}=\alpha_{\rm QED}/m$ from the residual electrons 
does not generate a birefringent signal to lowest order. 
A birefringent signal can be generated by quantum effects due to the electron spin or by coupling the 
electrons to the external magnetic field. 
However, these higher-order processes are suppressed in comparison the the lowest-order Thomson scattering 
${\mathfrak A}_{\rm T}=\alpha_{\rm QED}/m$.

For example, the birefringent signal from free electrons can be estimated from the Klein-Nishina formula 
\cite{KN1,KN2,Wig,Hei}
\bea
\label{Klein-Nishina} 
\frac{d\sigma_{\rm KN}^\perp}{d\Omega}
=
\frac{\alpha_{\rm QED}^2}{m^2}\,
\frac{\omega^2}{m^2}\,
\frac{\vartheta^4}{16}
\,,
\ea
where $\vartheta\ll1$ denotes the small scattering angle. 
The birefringent signal induced by coupling to the external magnetic field can be estimated by the 
insertion of an additional vertex into the lowest-order Feynman diagram for Compton scattering. 
The resulting amplitude is suppressed by the inverse combined Keldysh parameter \cite{ass1,ass2}  
\bea
\frac1\gamma=\frac{qB_{\rm ext}}{m\omega}
\,,
\ea
which is on the percent level for our parameters. 
Thus, these effects are sufficiently suppressed in comparison with the signal 
${\mathfrak A}_{\delta\Psi}^\perp$ considered here -- unless we have too many residual electrons. 

Even for a larger number of residual electrons, one could still observe the desired signal 
${\mathfrak A}_{\delta\Psi}^\perp$ if the unwanted contributions from the electrons would not 
add up coherently.
This would be quite natural since the light electrons are much more prone to de-phasing and decoherence 
than the heavy nuclei. 
They experience a stronger recoil and typically have a broader distribution in velocity and position.
Further, if the birefringent amplitude depends on the initial electron spin state of the (unpolarized) 
electrons or if their spin state is changed by the birefringent scattering event, one would get an 
incoherent superposition. 

%%%%%%%%%%%%%%%%%%%%%%%%%%%%%%%%%%%%%%%%%%%%%%%%%%%%%%%%%%%%%%%%%%%%%%%%%%%%%%%%%%%%%%%%%%%%%%%%%%%%%%%%%%%%%%%%%%
\subsection{Residual Radiation} 
%%%%%%%%%%%%%%%%%%%%%%%%%%%%%%%%%%%%%%%%%%%%%%%%%%%%%%%%%%%%%%%%%%%%%%%%%%%%%%%%%%%%%%%%%%%%%%%%%%%%%%%%%%%%%%%%%%
%
Further sources for x-ray photons which might lead to a false signal could be radiation processes, 
such as from the electrons accelerated by the optical laser during ionization or bremsstrahlung effects etc. 
However, these photons will typically have frequencies different from the XFEL frequency and thus 
can be filtered out efficiently by the very sharp energy resolution (in the sub-eV regime) of the 
x-ray polarization filters based on multiple Bragg scattering. 

%%%%%%%%%%%%%%%%%%%%%%%%%%%%%%%%%%%%%%%%%%%%%%%%%%%%%%%%%%%%%%%%%%%%%%%%%%%%%%%%%%%%%%%%%%%%%%%%%%%%%%%%%%%%%%%%%%
\subsection{Interferences} 
%%%%%%%%%%%%%%%%%%%%%%%%%%%%%%%%%%%%%%%%%%%%%%%%%%%%%%%%%%%%%%%%%%%%%%%%%%%%%%%%%%%%%%%%%%%%%%%%%%%%%%%%%%%%%%%%%%
%
Note that the amplitude ${\mathfrak A}_{\delta\Psi}$ derived here is imaginary. 
This is related to the parity $\mathcal P$ symmetry of QED as the Coulomb field ${\bf E}_{\rm ext}({\bf r})$, 
the symmetry-breaking tensor $\delta\Psi({\bf r})$, and thus also the effective current ${\bf J}_\omega^{\rm eff}$ 
generating ${\mathfrak A}_{\delta\Psi}$ are all odd functions of space. 
In contrast, $\delta\epsilon$ and $\delta\mu$ are even functions, such that the amplitude for pure 
Delbr\"uck scattering ${\mathfrak A}_{\rm D}$ is real -- as is the amplitude for Thomson scattering 
${\mathfrak A}_{\rm T}$. 
As a result, there are no interference terms (to leading order) between the field-assisted 
Delbr\"uck scattering ${\mathfrak A}_{\delta\Psi}$ considered here and Thomson scattering 
${\mathfrak A}_{\rm T}$ or Delbr\"uck scattering ${\mathfrak A}_{\rm D}$, while there can be 
interference terms between the latter two ${\mathfrak A}_{\rm T}$ and ${\mathfrak A}_{\rm D}$. 

%%%%%%%%%%%%%%%%%%%%%%%%%%%%%%%%%%%%%%%%%%%%%%%%%%%%%%%%%%%%%%%%%%%%%%%%%%%%%%%%%%%%%%%%%%%%%%%%%%%%%%%%%%%%%%%%%%
\section{Appendix: PIC simulations}%\label{comoving}
%%%%%%%%%%%%%%%%%%%%%%%%%%%%%%%%%%%%%%%%%%%%%%%%%%%%%%%%%%%%%%%%%%%%%%%%%%%%%%%%%%%%%%%%%%%%%%%%%%%%%%%%%%%%%%%%%%
The 3D particle-in-cell simulations shown in Fig. \ref{fig-pic} and supplementary Figs. \ref{3D-PIC} and \ref{fig::tweacInteraction} were performed using the code PIConGPU \cite{buss1,buss2}, version $0.4.3$ \cite{buss3} (Figs. \ref{fig-pic} and \ref{3D-PIC}) $0.4.2$ \cite{buss4} (Fig. \ref{fig::tweacInteraction}). For Figs. \ref{fig-pic} and \ref{3D-PIC}, the grid size is $640 \times 1280\times 640$, the spatial resolution is $256$ cells per $\lambda_0$ ($\lambda_0=800$ nm) and the temporal resolution is $6\times 10^{-3}$ fs. For Fig. \ref{fig::tweacInteraction} we employed a moving-window frame with a total size of $256 \times 2560 \times 1024$ cells which propagates for $50,000$ iterations. The spatial resolution here is $4.5/\lambda_0 \times 18/\lambda_0 \times 18/\lambda_0$ with a temporal resolution of $103$ as. The electromagnetic field evolution is simulated with the Yee solver \cite{yee}, while the particle motion is computed using the Boris pusher \cite{boris}. Particles influence the fields via the Esirkepov current deposition scheme \cite{esi} with a TSC macro-particle shape \cite{hock}. Ionization was taken into account for Figs. \ref{fig-pic} and \ref{3D-PIC} by field ionization via the tunneling and barrier suppression \cite{del} with orbital structure in shielding of the ion charge according to \cite{cle1,cle2}, as well as by collisions via a modified Thomas-Fermi model \cite{more}.

%%%%%%%%%%%%%%%%%%%%%%%%%%%%%%%%%%%%%%%%%%%%%%%%%%%%%%%%%%%%%%%%%%%%%%%%%%%%%%%%%%%%%%%%%%%%%%%%%%%%%%%%%%%%%%%%%%
\subsection{$50~\rm nm$ Carbon Cluster}%\label{comoving}
%%%%%%%%%%%%%%%%%%%%%%%%%%%%%%%%%%%%%%%%%%%%%%%%%%%%%%%%%%%%%%%%%%%%%%%%%%%%%%%%%%%%%%%%%%%%%%%%%%%%%%%%%%%%%%%%%%

\begin{figure}[h]
\includegraphics[width=.41\textwidth]{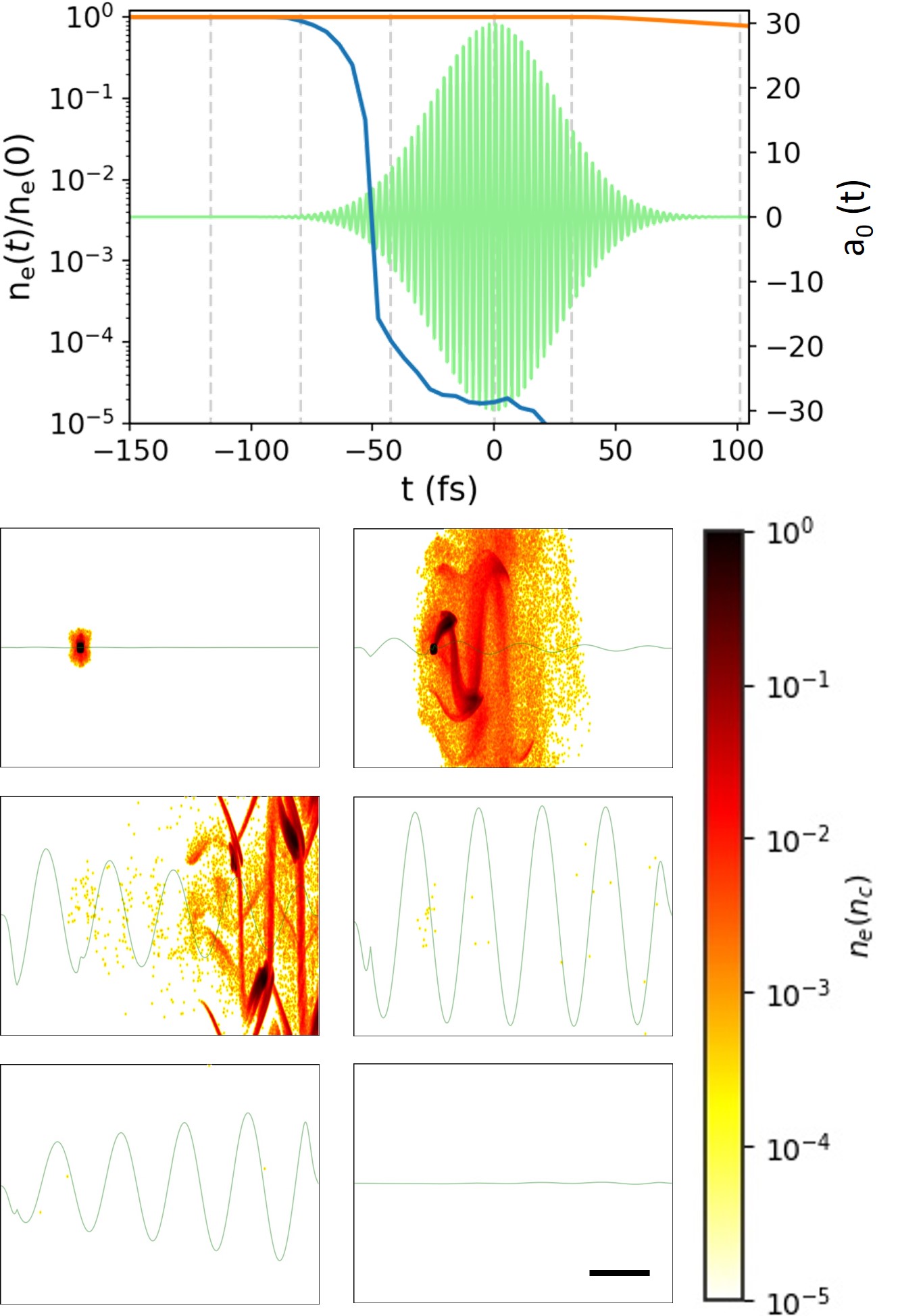}%
\caption{Same as in Fig.~\ref{fig-pic}, but for a cubic carbon cluster target at 50~nm side length.}
\label{3D-PIC}
\end{figure}

In analogy to Fig.~\ref{fig-pic}, the PIC simulation for a cubic carbon cluster target at 50~nm side length
is shown in Fig.~\ref{3D-PIC}.
For this smaller cluster, the number of electrons within the interaction volume drops by more than four 
orders of magnitude before the peak of the laser pulse is reached. 
In addition, the effect of the Coulomb explosion of the remaining ions is weaker (as expected). 

%%%%%%%%%%%%%%%%%%%%%%%%%%%%%%%%%%%%%%%%%%%%%%%%%%%%%%%%%%%%%%%%%%%%%%%%%%%%%%%%%%%%%%%%%%%%%%%%%%%%%%%%%%%%%%%%%%
%%%%%%%%%%%%%%%%%%%%%%%%%%%%%%%%%%%%%%%%%%%%%%%%%%%%%%%%%%%%%%%%%%%%%%%%%%%%%%%%%%%%%%%%%%%%%%%%%%%%%%%%%%%%%%%%%%

%%%%%%%%%%%%%%%%%%%%%%%%%%%%%%%%%%%%%%%%%%%%%%%%%%%%%%%%%%%%%%%%%%%%%%%%%%%%%%%%%%%%%%%%%%%%%%%%%%%%%%%%%%%%%%%%%%
\subsection{Co-moving Laser Focus}%\label{comoving}
%%%%%%%%%%%%%%%%%%%%%%%%%%%%%%%%%%%%%%%%%%%%%%%%%%%%%%%%%%%%%%%%%%%%%%%%%%%%%%%%%%%%%%%%%%%%%%%%%%%%%%%%%%%%%%%%%%

Laser-wakefield acceleration \cite{Esarey2009} can sustain a cavity with full electron blowout until reaching 
laser pump depletion. 
If an XFEL laser propagates in this blowout in the region for the entire duration where the laser is most 
intense, electron background is minimal, so that background processes are minimized.
For such an approach, lower densities of several $10^{19}~\rm cm^{-3}$ %$\SI{e+19}{\per\cm\cubed}$ 
are required, because the drive laser pulse dimensions have to be smaller than the plasma wavelength 
$2\pi c/\omega_p$.
Compared to high-density targets, see Figs.~\ref{fig-pic} and~\ref{3D-PIC}, 
this reduction in density is compensated by a much 
longer interaction length, which can be in the mm-range.
However for laser-wakefield acceleration, %LWFA 
the drive laser propagation needs to be parallel to XFEL propagation, greatly diminishing the cross-section 
of laser-assisted Delbr\"uck scattering, so that a different scattering geometry is chosen.

A simplified version of Traveling-wave electron acceleration (TWEAC) \cite{Debus2019}, i.e., using one drive 
laser only, provides a focal region co-moving at the vacuum speed of light with the XFEL laser. 
Locally in the interaction region of the XFEL laser, the ions %electron 
see an incoming plane wave coming at some incident angle $\phi$. 
Globally, the laser is cylindrically focused to a line collinear with the XFEL direction of propagation. 
Also, in order to maintain a stationary overlap with the XFEL beam, the laser pulse shape has to be 
prepared to feature a pulse-front tilt 
of $\phi/2$ \cite{Steiniger2019}.

The example shown in Fig.~\ref{fig::tweacInteraction} is based on a 3D-PIC simulation, displaying 
the electron density time-averaged over the full 1~mm %\SI{1}{mm} 
interaction length, showing a sizable region devoid of residual electrons 
(at least 3 orders of magnitudes electron density reduction) in the vicinity of the XFEL 
pulse and close to maximum laser intensity.

Neither the depletion limit, nor laser pulse guiding constraints limit aforementioned geometry and interaction length, 
but the laser pulse energy available. 
For intensities at several $10^{21}~\rm W/cm^2$ %$\SI{e+21}{W/cm^2}$ 
($a_0=30$), laser pulse energies only remain in a technically feasible range if the incidence 
angle does not become too large and features a tight focus. 
Furthermore, for minimal differences between initial onset of interaction and the later stationary 
interaction conditions, it is useful to choose an interaction region close to a steep edge of the gas target. 
This facilitates full plasma blowout around an XFEL laser over the entire interaction with a gas target.

\begin{widetext}

%%%%%%%%%%%%%%%%%%%%%%%%%%%%%%%%%%%%%%%%%%%%%%%%%%%%%%%%%%%%%%%%%%%%%%%%%%%%%%%%%%%%%%%%%%%%%%%%%%%%%%%%%%%%%%%%%%
\begin{figure}[h]
\includegraphics[width=0.8\textwidth]{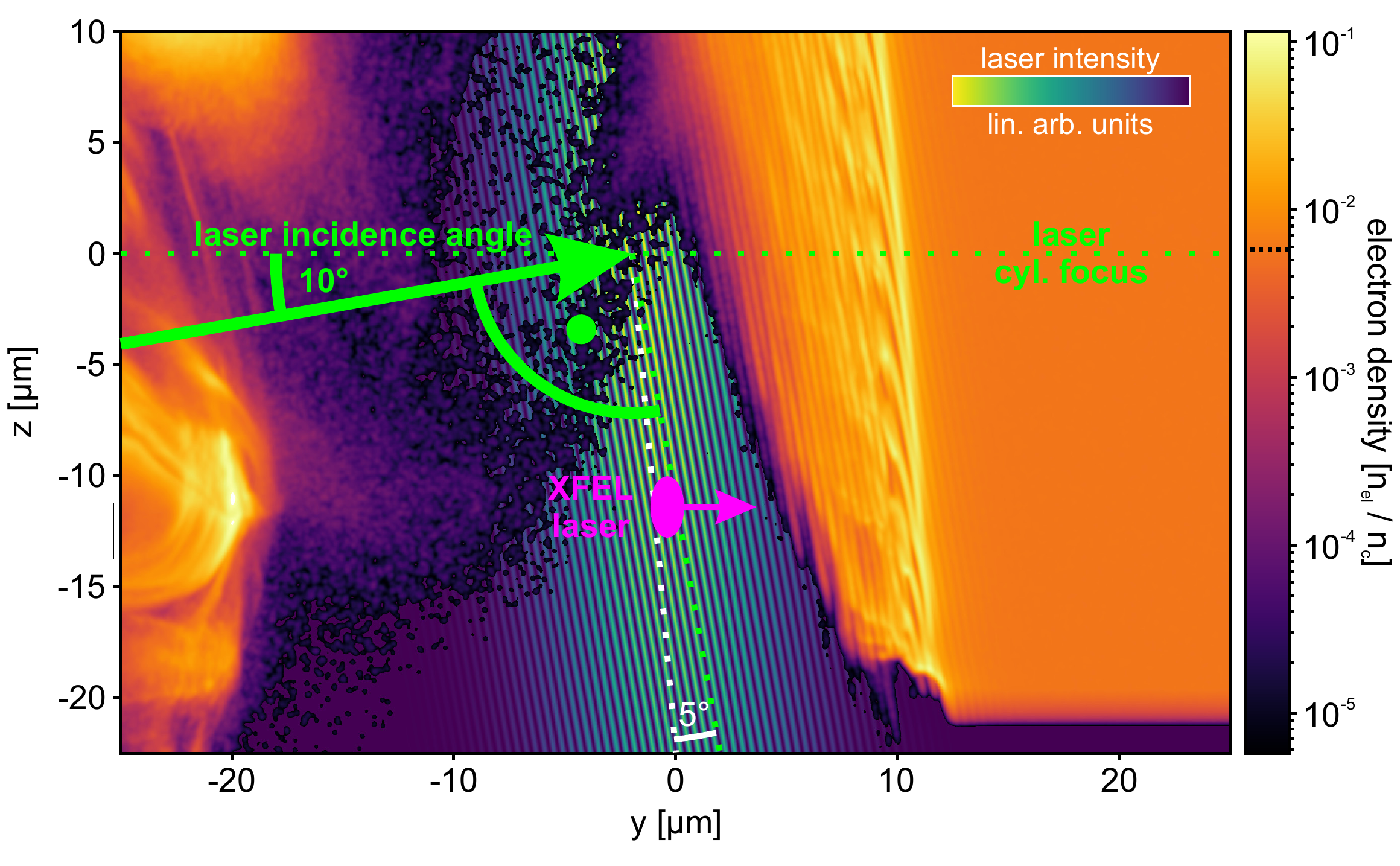}
\caption{In a 3D-PIC simulation (\emph{PIConGPU}), an electron blow-out cavity is driven in a 1~mm %\SI{1}{mm} 
long $H_2$ plasma target at $10^{19}~\rm cm^{-3}$ (black-dotted on density scale) %$\SI{1.0e+19}{\per\cm\cubed}$ 
using a cylindrically focused, pulse-front tilted laser 
(800~nm, 30~fs, 335~J, $1.9\times10^{21}~\rm W/cm^2$, $a_0=30$, 2.5~$\mu$m 
%\SI{800}{nm}, \SI{30}{fs}, \SI{335}{J}, $\SI{1.9e+21}{W/cm^2}$, $a_0=30$, \SI{2.5}{\um} 
focus height and 1~mm %\SI{1}{mm} 
interaction length) incident at $10^\circ$ %\SI{10}{\degree} 
with respect to the XFEL direction of propagation and $5^\circ$ pulse-front tilt.
For assessing the amount of residual electrons over the entire interaction length the electron density is 
time-averaged over the full 1~mm %\SI{1}{mm} 
interaction length, showing a sizable region cleared of electrons in the vicinity of the XFEL pulse 
and close to maximum laser intensity.}
\label{fig::tweacInteraction}
\end{figure}
%%%%%%%%%%%%%%%%%%%%%%%%%%%%%%%%%%%%%%%%%%%%%%%%%%%%%%%%%%%%%%%%%%%%%%%%%%%%%%%%%%%%%%%%%%%%%%%%%%%%%%%%%%%%%%%%%%

%\begin{thebibliography}{99}

%\bibitem{KN1}
%O.~Klein and Y.~Nishina, 
%{\it \"Uber die Streuung von Strahlung durch freie Elektronen nach der neuen relativistischen Quantendynamik von Dirac}, 
%Zeitschrift f\"ur Phys.\ {\bf 52}, 853 (1929).

%\bibitem{KN2}
%O.~Klein and Y.~Nishina,
%{\it The Scattering of Light by Free Electrons according to Dirac's New Relativistic Dynamics}, 
%Nature {\bf 122}, 398 (1928). 

%\bibitem{Wig}
%A.~Wightman, 
%{\it Note on polarization effects in Compton scattering}, 
%Phys.\ Rev.\ {\bf 74}, 1813 (1948).

%\bibitem{Hei}
%W.~Heitler, {\it The quantum theory of radiation}, Courier Corporation, 1984.

%\bibitem{ass1}
%R.~Sch\"utzhold, H.~Gies and G.~Dunne, 
%{\it Dynamically assisted Schwinger mechanism}, 
%Phys.\ Rev.\ Lett.\ {\bf 101}, 130404 (2008).

%\bibitem{ass2}
%G.~Torgrimsson, C.~Schneider and R.~Sch\"utzhold, 
%{\it Sauter-Schwinger pair creation dynamically assisted by a plane wave}, 
%Phys.\ Rev.\ D {\bf 97}, 096004 (2018).

%\end{thebibliography}

\end{widetext}


\begin{thebibliography}{99}
\bibitem{foot1}
One could derive the same result from QED Feynman diagrams such as in Fig.~\ref{figure}. 
%
Since three vertices (the two XFEL photons and the external magnetic field) correspond to low 
energies and momenta (i.e., well below the electron mass $m$), energy-momentum conservation implies 
that the fourth vertex (representing the Coulomb field) does also involve low energies and momenta.
%
In fact, for the case of forward scattering considered here, these energies and momenta are extremely low. 
%
Thus, we may use the low-energy limit of the QED Feynman diagrams, which are equivalent to the 
Euler-Heisenberg Lagrangian~\eqref{EHL}. 
%
Note that the situation is different for pure Delbr\"uck scattering where only two vertices 
(the two XFEL photons) correspond to low energies and momenta such that the other two 
(representing the Coulomb field) can involve high momenta.
\bibitem{foot2}
This estimate can also be obtained from the lowest-order Feynman diagram. The four vertices of the fermion box yield $q^4$ and the two external 
Coulomb field lines contribute $(Zq)^2$. Furthermore, scaling everything with respect to the electron mass 
(as the natural QED scale), the two external XFEL photon lines scale 
with $\omega^2/m^2$.
\end{thebibliography}
\end{document}